# The spacetime metric in particle QED

Charles Francis

ABSTRACT
A small time delay between interactions, which has previously been shown to remove divergences from QED, is used to show that, if spacetime geometry is emergent from particle interactions in the manner suggested by Bondi, then Minkowski metric, which appears in the Schrödinger equation as a requirement of the probability interpretation, is perturbed in physical measurement, leading to curved spacetime in accordance with Einstein's equation.

## 1 Conceptual background

Sir Arthur Stanley Eddington (1923) expressed the view that *"A physical quantity is defined by the series of operations and calculations of which it is the result"*. Although he was writing in the context of the general theory of relativity, this view acquires support from quantum theory in which it is observed that "*In the general case we cannot speak of an observable having a value for a particular state, but we can ... speak of the probability of its having a specified value for the state, meaning the probability of this specified value being obtained when one makes a measurement of the observable"* (Dirac, 1958). According to such a view, the mathematical structure of spacetime does not exist an ontological prior, an arena into which matter is placed in the manner of Newtonian absolute space, but is instead an emergent property arising from the interactions of matter (and energy) with matter (and energy).

Sir Hermann Bondi adopted just such a position in his common sense treatment of relativity (1964). By basing the definition of spacetime coordinates on the radar method, and noting that *"... the length of a rigid rod is determined by the electrical interactions of the atoms and, therefore, in fact, by a superposition of radar methods"*, Bondi came to the conclusion *"So, with our modern outlook and modern technology the Michelson-Morley experiment is a mere tautology"* (Bondi 1967). Bondi's conclusion does not depend upon the specific choice of radar for the definition of coordinates, since the general theory of relativity is coordinate independent. It does depend on the recognition, provided by the special theory of relativity, that the definition of the metre based on the constancy of the speed of light is consistent with any other useful definition of the metre which might have been chosen, and on the fundamental dependency of the structures of matter, including their metric properties, on the electromagnetic force.

Francis (2013) has given a construction of particle theoretical QED from relativistic quantum mechanics and avoiding divergences from loop diagrams, from the Dyson instability and from the Landau pole. In this construction a particle is a pointlike, or size-less, entity for which position is, in general, indeterminate because there is no background space. The quantum state (including the wave function) is a mathematical device describing knowledge not ontology, and never perfectly describes a point because it is not possible to have complete knowledge of a physical system governed by indeterminist laws at a fundamental level. Fields are operators used in the description of interactions between particles. It can be seen in this construction that, although position states space are defined on $\mathbb{R}^4$, the spacetime metric does not appear at a fundamental level of the physical theory. The (relativistic) Schrödinger equation is shown from the requirements of the probability interpretation by way of Stone's theorem (1932) (appendix A). To find the general solution of the Schrödinger equation, we take the Fourier transform of the initial wave function, which introduces Euclidean metric into the mathematical description. Then the evolution of momentum eigenstates generates Minkowski metric in the exponent of the free particle solution. It is thus seen that Minkowski metric appears naturally in the evolution of quantum states, and is a property of a mathematical configuration space, rather than a physical property of prior spacetime.

A question arises because the implication is that the Schrödinger equation must necessarily be solved in Minkowski spacetime. This is resolved in Francis (2014) because the wave function is understood not as an ontological entity but as a mathematical construct and evolves not in curved spacetime but in a non-physical affine Minkowski spacetime. States are projected back to physical spacetime in measurement, when the wave function collapses. This formulation of quantum mechanics was shown to be consistent with general relativity because gravitational redshifts for quantum states are identical to gravitational redshifts for classical electromagnetic radiation. Geodesic motion was found for classical matter, consisting of many quantum motions, and physical spacetime is found from the envelope of configuration spaces.

Since the spacetime metric is not assumed as fundamental, it is required to show how it emerges in the physical behaviour of matter. It will be seen here that the introduction particle interactions, as required for a theory of physical measurement and for a description all structures of matter in our immediate environment, causes a small perturbation to Minkowski metric such that massive elementary particles generate curvature equivalent to that given by Einstein's equation. In standard treatments of general relativity, Einstein's equation is determined because it is a second rank tensor equation relating curvature (defined only from the metric) with energy-momentum and such that the covariant law of conservation of energy-momentum is automatically satisfied by obedience of the Einstein tensor to the contracted Bianchi identity. The argument of section 2 for the gravity of an elementary particle does not replace this derivation, but rather shows how metric properties derived from Einstein's equation can be directly related to a discrete time interval between interactions of particles in QED.

## 2 A Modification to Radar

Although Bondi did not delve into the intricacies of photon exchange in QED, he did make the point that the structures of matter in our immediate environment are determined by the electromagnetic force (to which one should add the Pauli exclusion principle). In a particle theoretic treatment of QED, the electromagnetic force is seen as resulting on the exchange of photons, which is in essence the process used by the radar method of determining distance. Bondi's clear implication is that the metric in special relativity actually arises from these electromagnetic processes taking place in at a fundamental level in the structure of matter. If this is correct it should be the case that a more careful treatment should also explain curvature.

Consistent with Einstein's 1905 paper and the internationally agreed empirical definition of the metre, Bondi's $k$-calculus for special relativity postulates instantaneous reflection of radar at the event whose position is to be determined. Although reflection clearly takes place on a very small time scale, there is no empirical basis on which we can say it is actually instantaneous. A natural generalization is to hypothesis a small time delay between absorption and emission in proper time of a fundamental charged particle (electron or quark) reflecting electromagnetic radiation. There are several reasons for introducing such a delay.

Firstly, as shown here, the delay perturbs the metric; if geometry emerges from particle interactions and if the reflection of a photon were instantaneous, the physical metric would be Minkowski, but a small delay in reflection leads to a source of curvature at the event where the reflection takes place. The argument below will relate this to Einstein's equation by showing that a single isolated particle with an inherent delay in reflection generates Schwarzschild geometry.

Secondly, it is well known that some small-scale modification is needed to qed in order to remove the ultraviolet divergence in loop diagrams, the Landau pole, and the Dyson instability. A small time delay is an effective cut-off and was used by Francis (2013) to allow the rigorous construction of a consistent QED.

Finally, a minimum time between interactions proportional to mass may be related to the concept of inertia; if the interactions of muons and electrons with photons are discrete and identical, then it is natural that the acceleration due to the electromagnetic field will be proportional to the frequency of interaction. Consequently, an intrinsic delay between interactions proportional to mass will result in accelerations inversely proportional to mass.

An intrinsic delay between the interactions of elementary particles affects the empirical definition of spacetime measurement (e.g., SI units). We seek to analyse the geometric implications. The metric is determined as in the $k$-calculus for special relativity, from the minimum time for the return of information reflected at an event. But now this minimum net time depends not only on the maximum theoretical speed of information, $c$, but also on an effective least proper time between absorption and emission in the reflection of a photon. Let the effective time delay be $4Gm$, where $m$ is the mass of the reflecting particle and $4G$ is a constant of proportionality (it will be found that $G$ is the gravitational constant). Special relativity is recovered in the limit in which $G$ goes to zero (allowing $G$ to go to zero re-introduces the Landau pole and the Dyson instability so this limit may not be valid).

Consider a system with a single gravitating elementary particle in an eigenstate of position at O. Position eigenstates span Hilbert space and will be sufficient for a description of geometry. Gravity from other sources is ignored. The Schwarzschild metric has the form

$$\mathrm{d}s^2 = k^2\mathrm{d}t^2 - k^{-2}\mathrm{d}r^2 - r^2(\mathrm{d}\theta^2 + \sin^2\theta\,\mathrm{d}\phi^2)\,, \tag{2.1}$$

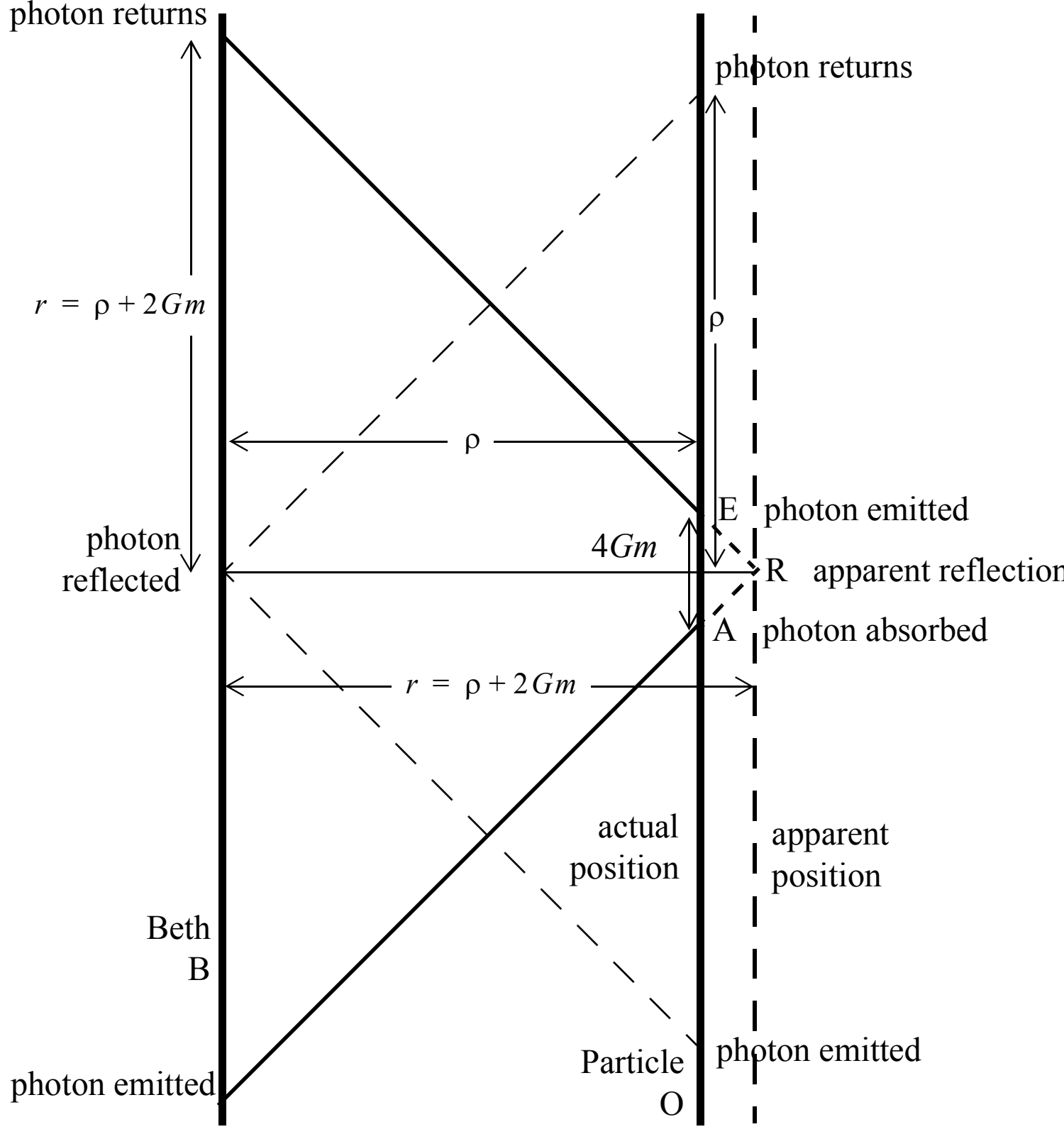

**Figure 1:** Beth's determination of coordinates of a gravitating particle seen in configuration space at particle position O.

found by solving Einstein's equation. The metric is everywhere defined by physical processes equivalent to the reflection of photons, as described in the radar method, where the reflection is treated as instantaneous, as in Bondi's *k*-calculus for special relativity, except for the particle at O, where it is assumed that there is a small inherent time delay $4Gm$ between absorption and emission of the reflected photon. This idealisation will enable us to see that the effect of a small time delay in the interactions for a single particle is equivalent to gravity. Clearly the reality is that all particles gravitate. It may thus be thought that charged particles generate curvature according to the mechanism described here, and that more generally consistency determines that mass energy generates curvature in order to satisfy the contracted Bianchi identity.

As described in Francis (2014), quantum motions are described in tangent space. Consequently proper time for the gravitating particle is equal to time at infinity. An isolated elementary particle in an eigenstate of position has spherical symmetry so that spacetime diagrams may be used to show the radial coordinate in $n$ dimensions without loss of generality. An observer, Beth, at B, measures the position, O, of the particle by the radar method. Figure 1 is a spacetime diagram showing a tangent space at O in coordinates with unit speed of light, so that light is shown at 45°, lines of equal time are horizontal and time is proper time for the gravitating particle. Tangent space at the particle is defined ignoring curvature caused by the particle itself, and is more correctly called configuration space. The coordinate distance between Beth and the particle is $\rho$. The reflection of the photon is not instantaneous, but is seen as separate absorption and emission events, A & E, so that the distance, $r$, of the apparent position of reflection, R, as determined by the radar method is greater that the value, $\rho$, in Minkowski configuration space which would be determined if the reflection were truly instantaneous and Beth's clock time were equal to particle proper time. It should be noted that because the definition of distances depends upon physical processes, $r$ is an actual distance as determined by Beth (we distinguish configuration space distances defined by a particular observer from proper distances, found by integrating local distances along a path). If the effective delay in the reflection is $4Gm$, then the coordinate distance of R from Beth is

$$r = \rho + 2Gm \tag{2.2}$$

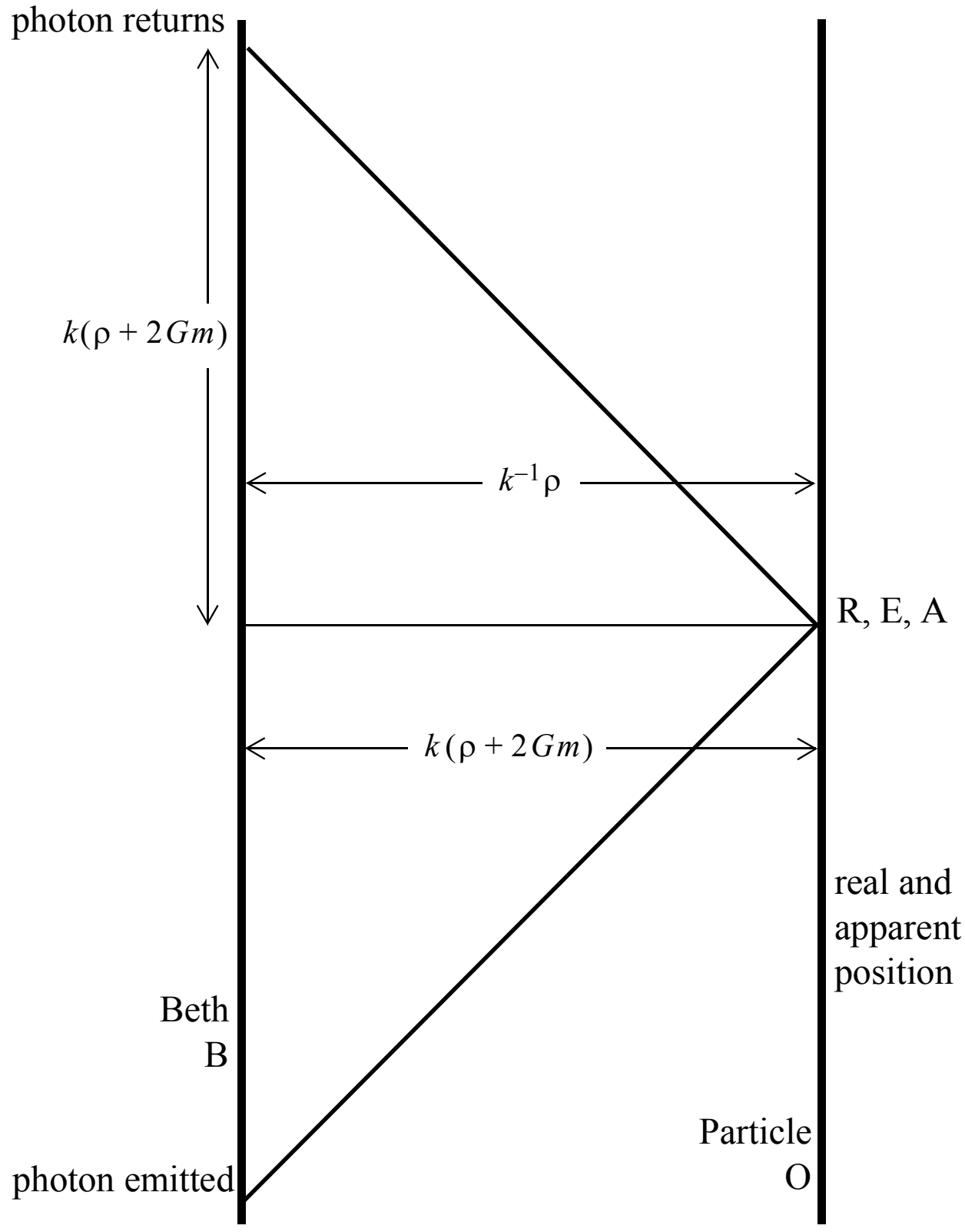

**Figure 2:** Beth's determination of coordinates of a gravitating particle seen in configuration space at Beth's position B. Beth cannot resolve a difference of position between absorption and re-emission of the photon at R.

Figure 2 shows Beth's measurement in configuration space at Beth's position, B. For simplicity we describe configuration space ignoring the non-relativist orbital velocity (or proper acceleration) of Beth, required so that she remains at constant distance from the particle. Because the coordinates are those determined from measurement, Beth cannot resolve the difference of position between A, E, and R. Since the transmission of photons is described in configuration space, which is flat, the scaling factors in the metric (2.1) apply globally. Hence, the time on Beth's clock is $k(\rho + 2Gm)$ and the distance is $k^{-1}\rho$. According to the radar method we require

$$k(\rho + 2Gm) = k^{-1}\rho . \tag{2.3}$$

So,

$$k^2 = \frac{\rho}{\rho + 2Gm} . \tag{2.4}$$

Substituting (2.2)

$$k^2 = \frac{r - 2Gm}{r} = 1 - \frac{2Gm}{r} . \tag{2.5}$$

Thus the Schwarzschild metric is found in coordinates using proper time for the gravitating particle and distance, $r$, determined by the radar method remotely from the gravitating particle.

## 3 Event horizon of a point particle

It is sometimes claimed that the concept of a massive point particle is inconsistent with classical general relativity. In particular, proper time would stop at its Schwarzschild radius, so that interactions with other matter would be prohibited. The argument above resolves this paradox because the metric is not fundamental. The description of a particle as a point (or point-like) simply means that it has no size. According to (2.2), the point $\rho = 0$ in configuration space at O maps to $r = 2Gm$ in Schwarzschild coordinates. Transmission from the event horizon is not impossible, because quantum mechanics requires that the wave function is solved in flat configuration space (Francis 2014).

Thus, in Schwarzschild coordinates the particle has radius $r = 2Gm$ (where radius means radial coordinate). The region $r < 2Gm$ has no physical meaning. Only space outside the event horizon is mapped in these coordinates. Thus a point particle is mapped to the event horizon in a configuration space at the position of a distant observer, Beth. The interpretation of the metric at the Schwarzschild radius is not that time stops, but rather that coordinates are defined such that a finite proper time (the time of reflection) is describe in coordinates in which it has a zero value. A discrete interval of proper time between interactions is mathematically, but not physically, equivalent to a coordinate singularity at $r = 2Gm$ which "magnifies" a point particle at $\rho = 0$ to the *apparent* size of the event horizon.

It is also argued that the event horizon cannot be a point because the surface area of a sphere surrounding the particle is finite in the limit as the radius of the sphere tends to zero, suggesting that the event horizon itself must have a finite surface area. Since a point cannot have finite surface area, it is thought that the event horizon is not a point. This argument fails for two reasons. Firstly, it does not show that a point has finite surface area, but that the metric has a discontinuity. Secondly, a sphere is an object in classical geometry, which applies in good approximation on large scales, but does not make sense in a quantum treatment of single particles in which there is no background spacetime. It is meaningless to talk of the surface area of sphere surrounding the particle in the limit in which the radius of the sphere tends to zero because this assumes a substantive space in which such a sphere can be drawn.

The treatment of section 2 does not depend on spin. Spin is not classical angular momentum and does not generate a Kerr metric or Kerr-Newman metric, which apply when there is classical rotating matter involving many particles. The Reissner-Nordström metric is found by including the field of a charged mass in the stress energy tensor. It has been argued that the Reissner-Nordström metric is not consistent with the idea of an electron as a point particle because it has event horizons at

$$r_{\pm} = \frac{1}{2}(r_S \pm \sqrt{r_S^2 - 4r_Q^2}), \tag{3.1}$$

where $\rho_S = 2GM$ is the Schwarzschild radius, and $r_Q = Q^2G/4\pi\varepsilon_0$ where $Q$ is charge and $\varepsilon_0$ is the permittivity of the vacuum. For an electron, $r_S = 1.353\times10^{-57}$m, $r_Q = 9.152\times10^{-37}$m. Then (3.1) has no solution. However, since $r_{\pm} < r_S = 2\text{Gm}$ (when $r_{\pm}$ is real), the Reissner-Nordström metric has no coordinate singularity at, or outside, the Schwarzschild radius and the argument of section 2 is not affected. This should be expected because the gravity of the electromagnetic field is due to photons *outside* the Schwarzschild radius, and the region inside the Schwarzschild radius has no physical meaning.

## 4 Black holes

This argument has direct implications for both the physical structure and the behaviour of black holes. It is usually argued that there is a removable singularity at the event horizon, and that the general principle of relativity implies that the local structure of spacetime at the Schwarzschild radius is not intrinsically different from spacetime structure at any other point (e.g., Misner, Thorne, Wheeler 1973). Consequently it is thought possible to extend spacetime through the event horizon. In this case, the general principle must eventually break down at a singularity at the centre of the black hole. This mathematical argument assumes a classical structure for spacetime inconsistent with quantum theory and does not apply if the event horizon is seen as the point where interactions actually take place, as in figure 1.

In an idealised model of a black hole, the single gravitating particle in figure 1 is replaced by a large number of elementary particles occupying the same position in space (possible in principle for point particles). The interaction time for each particle is slowed by the gravity of the other particles relative to the time coordinate at infinity. In coordinates defined with constant radial velocity of light, the event horizon of a black hole of mass $M$ is again described coordinates with unit light speed, and light can, in principle, be reflected from the event horizon of a black hole after a time delay $4GM$ (figure 3). This is possible because quantum motions are described in flat configuration space, not by geodesic motion on curved spacetime. It does not violate the general principle of relativity, because in all cases interactions take place at the Schwarzschild radius. It may not be strictly necessary for the mass to be at a point, but it must be contained within the event horizon as viewed in configuration space at O. In practice gravitational collapse will ensure that the mass is contained in a very small region.

It remains possible to describe Hawking radiation arising from spontaneous pair creation, with the infall of the antimatter particle into the black hole, but a greater source of radiation may be described by relativistic quantum tunnelling. Wave functions for particles are plane waves in configuration space, and can be emitted to infinity provided that there is sufficient energy in the initial state. Matter in the hole will have high energy from gravitational collapse, and in addition, as the star collapses to a point at $\rho = 0$, the uncertainty principle requires wave functions with components with indefinitely high energies. We may conclude that localisation of matter at $\rho = 0$ creates energy states from which particles are radiated with relativistic velocities. Thus, relativistic quantum tunneling means that the mass of a black hole cannot actually be confined at a point.

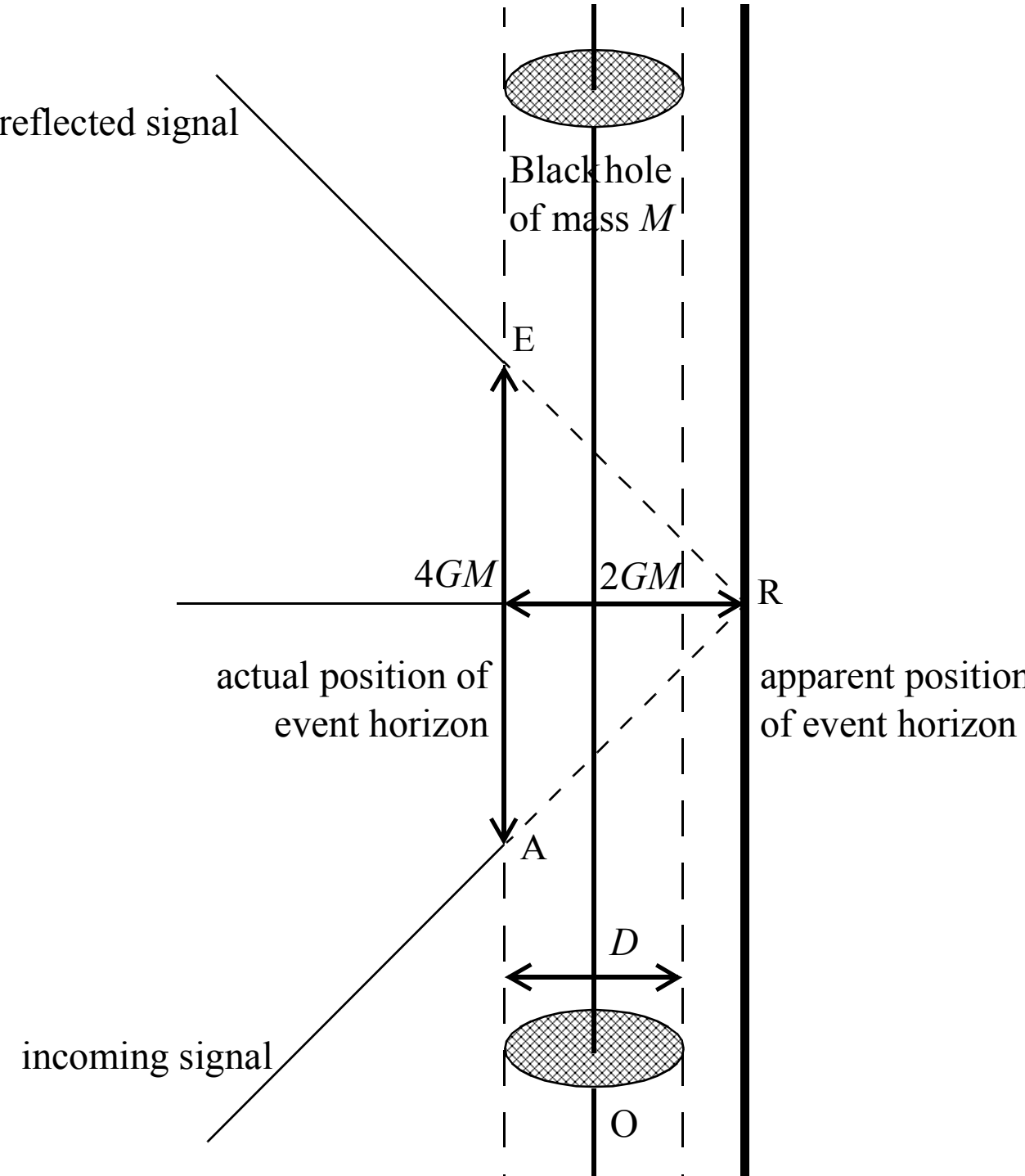

**Figure 3:** Idealised particle theoretic model of a black hole at O, using time coordinate defined from a clock at infinity and unit light speed. It does appear strictly necessary that the mass is contained at a single point in configuration space, but it must be contained in a small region.

It is to be expected that the greater the mass of the black hole and the quantity of infalling matter, the greater the amplitudes of states of sufficient energy to be radiated to infinity, and the greater the consequent radiation. This can be considered as a candidate for the mechanism underlying radiation by quasars in the early universe. Black holes in galactic cores in the early universe can be expected to have had an irregular structure and large amounts of infalling matter, so that radiation took place in all directions. Since matter is freely radiated from states with high energies, in the absence of further infalling matter, the black hole will rapidly cool and the quantity of radiation will reduce. As the black hole begins to stabilise, angular momentum of infalling matter will generate a disc. The direction of radiation will tend to lie along the axis of rotation, suggesting a mechanism for relativistic jets emitted from the core of many galaxies. Infalling matter will trigger further radiation. A possible explanation for gamma ray bursts is that they result when stars fall into a black hole, causing a sudden increase in radiated energy.

## 5 Pre-expansion as an ametric phase

If spacetime is not prior, and the metric is emergent from particle interactions, then in the neighbourhood of any singularity in classical general relativity there must be a region such that a classical metric structure does not apply. At the initial singularity, all particles were effectively in the same place and relative position had no meaning. It was then not possible to divide the early universe into indefinitely small regions which did not communicate and the horizon problem cannot be formulated. It is not, then, necessary to postulate inflation. Rather than rapid inflation from a small size, there was an initial state for which we cannot talk of spatial dimension or size, and when horizons did not exist. It would have taken many discrete intervals of proper time for interactions between elementary particles to establish the properties of a Riemannian manifold. Prior to that the image is one of perfect chaos, in which any photon may interact with any charged particle, so that the entire is causally connected. Perfect chaos in physical conditions gives rise to perfect order in a probabilistic description as required by quantum theory. Because positions cannot be distinguished, an ametric phase can only generate an isotropic initial condition for normal expansion.

It does not appear necessary to postulate that all matter participated in the creation of spacetime. Indeed, if some matter remained disconnected from the observable universe when spacetime formed, it could account for the observed matter/antimatter imbalance without the need to postulate an exotic and unobserved process in particle physics such as proton decay.

A lower bound for the duration of the ametric phase can be estimated by applying a Doppler shift to one interval of discrete time as appropriate to the high energies of particles near the big bang. Typical quoted energies for particles near the big bang are in the order of a factor $10^{30}$ greater than rest mass. In this case the discrete interval of proper time $10^{-65}$s for an electron is redshifted to $10^{-35}$s, within range of the time scales normally postulated for the end of inflation and the beginning of normal expansion.

## 6 Conclusion

A small inherent time delay between interactions was found by Francis (2013) to regularise QED and remove the Landau pole and the Dyson instability. A delay proportional to mass is directly related to inertia, since more frequent interactions with photons will lead to proportionately greater response. By considering the physical metric as consequent on the ontological process of photon exchange it has been seen that this small time delay requires that Minkowski metric is replaced with a metric obeying Einstein's equation, where the delay is proportional to mass. The argument describes a point particle at $\rho = r - 2Gm = 0$, where $r$ is the Schwarzschild radial coordinate and $2Gm$ is the Schwarzschild radius and is equal to half the inherent minimum time between interactions in particle QED. The exterior region of a Schwarzschild geometry is found to be equivalent to using the radar method to determine metric distances when this small time delay in reflection is taken into account.

Space has no meaning in Feynman diagrams. It was seen in Francis (2013) that it emerges in the macroscopic properties of particle QED, in which the underlying physical structures are described mathematically as graphs. Conservation of energy and momentum was derived from the integral formulae for probability amplitudes associated with Feynman diagrams, and follows from the principle that the interactions of elementary particles are always and everywhere the same. The view that space is an emergent property of the interactions of particles casts a deeper light on Einstein's equation. Instead of the duality between matter and space described by Wheeler *"Matter tells space how to curve. Space tells matter how to move",* at a fundamental level we have only matter. Spacetime describes relationships in matter, constrained by conservation of energy-momentum, as described by Einstein's equation.

Figure 1, from which Schwarzschild geometry surrounding a single gravitating particle is found, can be embedded into any geometry satisfying Einstein's equation. Then it is seen that a charged particle induces Einstein curvature equal to stress energy for that particle. This applies for charged particles, but Einstein's equation follows generally from the contracted Bianchi identity together with conservation of energy momentum, as required by the locality condition in QED. It is thus seen that the underlying cause of curvature is an effective small time delay in the interactions of elementary particles.

The surface area of a point particle is not physically meaningful, and nor does it make physical sense to extend the coordinate system to the interior of a particle. This is consistent with quantum theory in which it is in general not meaningful to talk of physical quantities which are not physically measured.

When the model is applied to an aggregate of many particles, each charged particle is it is expected to act as a source of curvature satisfying Einstein's equation. Einstein's equation is then also required for uncharged particles by covariant energy conservation and the contracted Bianchi identity. When the model is applied to a black hole, it is found that quantum interactions can take place at the Schwarzschild radius, and that energy and matter can be emitted by relativistic quantum tunnelling. Relativistic quantum tunnelling can potentially emit large quantities of energy from black holes, and should be considered as a candidate mechanism for quasars, gamma ray bursts, and the relativistic jets emitted from the cores of many galaxies. An ametric phase offers an alternative to inflation as a possible resolution of the horizon problem.

## Appendix A Justification of the Schrödinger equation

If the ket at time $t_0$ was either $|f(t_0)\rangle$ or $|g(t_0)\rangle$, then it will evolve into either $|f(t)\rangle$ or $|g(t)\rangle$ at time $t$. This requires that $U$ is linear

$$U(t, t_0)(a|f(t_0)\rangle + b|g(t_0)\rangle) = aU(t, t_0)|f(t_0)\rangle + bU(t, t_0)|g(t_0)\rangle . \tag{A.1}$$

Since local laws of physics are always the same, and $U$ does not depend on the ket on which it acts, the form of the evolution operator for a time span $t$, $U(t) = U(t+t_0, t_0)$ does not depend on $t_0$. We require that the evolution in a span $t_1 + t_2$ is the same as the evolution in $t_1$ followed by the evolution in $t_2$, and is also equal to the evolution in $t_2$ followed by the evolution in $t_1$, $U(t_2)U(t_1) = U(t_2+t_1) = U(t_1)U(t_2)$. In zero time span, there is no evolution. So, $U(0)$ does not change the ket; $U(0) = 1$. Using negative $t$ reverses time evolution (put $t = t_1 = -t_2$); $U(-t) = U(t)^{-1}$.

In the absence of further information, the result of the calculation of probability of a measurement result $g$ at time $t_2$ given an initial condition $f$ at time $t_1$ is not affected by the time at which it is calculated. Since kets can be chosen to be normalised we may require that $U$ conserves the norm, i.e., for all $|g\rangle$,

$$\langle g|U^\dagger U|g\rangle = \langle g|g\rangle. \tag{A.2}$$

Applying (A.2) to $|g\rangle + |f\rangle$,

$$(\langle g| + \langle f|)U^\dagger U(|g\rangle + |f\rangle) = (\langle g| + \langle f|)(|g\rangle + |f\rangle). \tag{A.3}$$

By linearity of $U$,

$$(\langle g|U^\dagger + \langle f|U^\dagger)(U|g\rangle + U|f\rangle) = (\langle g| + \langle f|)(|g\rangle + |f\rangle). \tag{A.4}$$

By linearity of the inner product,

$$\begin{aligned}\langle g|U^\dagger U|g\rangle + \langle g|U^\dagger U|f\rangle + \langle f|U^\dagger U|g\rangle + \langle f|U^\dagger U|f\rangle \\ = \langle g|g\rangle + \langle g|f\rangle + \langle f|g\rangle + \langle f|f\rangle\end{aligned} \tag{A.5}$$

Thus, from (A.2),

$$\langle g|U^\dagger U|f\rangle + \langle f|U^\dagger U|g\rangle = \langle g|f\rangle + \langle f|g\rangle. \tag{A.6}$$

Similarly conservation of the norm of $|g\rangle + i|f\rangle$ gives

$$\langle g|U^\dagger U|f\rangle - \langle f|U^\dagger U|g\rangle = \langle g|f\rangle - \langle f|g\rangle. \tag{A.7}$$

Combining (A.6) and (A.7) shows that $U$ is unitary, i.e. for all $|f\rangle, |g\rangle \in \mathbb{H}$,

$$\langle g|U^\dagger U|f\rangle = \langle g|f\rangle. \tag{A.8}$$

**Theorem:** (Marshall Stone, 1932) Let $\{U(t): t \in \mathbb{R}\}$ be a set of unitary operators on a Hilbert space, $\mathbb{H}$, $U(t): \mathbb{H} \to \mathbb{H}$, such that $U(t+s) = U(t)U(s)$ and

$$\forall t_0 \in \mathbb{R}, |f\rangle \in \mathbb{H}, \lim_{t \to t_0} U_t|f\rangle = U_{t_0}|f\rangle \text{ (strong continuity)}$$

then there exists a unique self-adjoint operator $H$ such that $U(t) = e^{-iHt}$.

*Proof:* The derivative of $U$ is

$$\begin{aligned}\dot{U} = \frac{dU}{dt} &= \lim_{dt \to 0} \frac{U(t+dt) - U(t)}{dt} = \lim_{dt \to 0} \frac{U(dt)U(t) - U(t)}{dt} \\ &= \left(\lim_{dt \to 0} \frac{U(dt) - 1}{dt}\right) U(t) = U(t)\left(\lim_{dt \to 0} \frac{U(dt) - 1}{dt}\right).\end{aligned} \tag{A.9}$$

This prompts the definition of the Hamiltonian operator:

**Definition:** The **Hamiltonian** $H$: $\mathbb{H} \to \mathbb{H}$ is given by

$$H = i\left(\lim_{dt \to 0} \frac{U(dt) - 1}{dt}\right). \quad (A.10)$$

The Hamiltonian has no dependency on $t$. We have

$$\dot{U}(t) = -iHU(t) = -iU(t)H. \quad (A.11)$$

So $-iH = U^{\dagger}\dot{U} = \dot{U}U^{\dagger}$. Since $U$ is unitary, for a small time $dt$,

$$1 = U^{\dagger}(t + dt)U(t + dt) \approx [U^{\dagger}(t) + \dot{U}^{\dagger}(t)dt][U(t) + \dot{U}(t)dt]. \quad (A.12)$$

Ignoring terms in squares of $dt$, and using $-iH = U^{\dagger}\dot{U}$, $iH^{\dagger} = \dot{U}^{\dagger}U$,

$$U^{\dagger}(t)U(t) - iH^{\dagger}dt + iHdt \approx 1.$$

Using unitarity of $U$, we find that $H$ is Hermitian, $H = H^{\dagger}$. (A.11) has solution

$$U(t) = e^{-iHt}. \quad (A.13)$$

**Corollary:** The wave function satisfies the Schrödinger equation

$$\partial_0 f(t, x) = -iHf(t, x). \quad (A.14)$$

*Proof:* Differentiate the wave function using (A.11),

$$\partial_0 f(t, x) = \langle x|\dot{U}|f(0)\rangle = \langle x|-iHU(t)|f(0)\rangle = \langle x|-iH|f(t)\rangle = -iHf(t, x). \quad (A.15)$$

**Corollary:** Newton's first law.

*Proof:* For a non-interacting particle, $H = E = \text{const}$ where $E^2 = (p^0)^2 = m^2 + \boldsymbol{p}^2$ for some constant $m$. For an initial state $|f\rangle$ with momentum space wave function $\langle p|f\rangle$, the general solution is

$$f(x) = \left(\frac{1}{2\pi}\right)^{3/2}\int d^3p\langle p|f\rangle\, e^{-ix \cdot p} \quad (A.16)$$

Thus the momentum space wave function $\langle p|f\rangle$ does not change in time. It should be noted that the dot product, $x{\cdot}p$, in (A.16) uses Minkowski metric. It arises from the solution of a differential equation, *not* the physical metric of spacetime.